# Non-Fermi surface nesting driven commensurate magnetic ordering in Fe-doped $Sr_2RuO_4$

M. Zhu[1], K. V. Shanavas[2], Y. Wang[3], T. Zou[1], W. F. Sun[3], W. Tian[4], V. O. Garlea[4], A. Podlesnyak[4], M. Matsuda[4], M. B. Stone[4], D. Keavney[5], Z. Q. Mao[3], D. J. Singh[2], and X. Ke[1]*

[1]*Department of Physics and Astronomy, Michigan State University, East Lansing, Michigan 48824, USA*

[2]*Department of Physics and Astronomy, University of Missouri, Columbia, Missouri 65211-7010, USA*

[3]*Department of Physics and Engineering Physics, Tulane University, New Orleans, Louisiana 70118, USA*

[4]*Quantum Condensed Matter Division, Oak Ridge National Laboratory, Oak Ridge, Tennessee 37831, USA*

[5]*Advanced Photon Source, Argonne National Laboratory, Argonne, Illinois 60439, USA*

*Corresponding author: ke@pa.msu.edu.

$Sr_2RuO_4$, an unconventional superconductor, is known to possess an incommensurate spin density wave instability driven by Fermi surface nesting. Here we report a static spin density wave ordering with a commensurate propagation vector $\mathbf{q_c} = (0.25\ 0.25\ 0)$ in Fe-doped $Sr_2RuO_4$, despite the magnetic fluctuations persisting at the incommensurate wave vectors $\mathbf{q_{ic}} = (0.3\ 0.3\ L)$ as in the parent compound. The latter feature is corroborated by the first principles calculations, which show that Fe substitution barely changes the nesting vector of the Fermi surface. These results suggest that in addition to the known incommensurate magnetic instability, $Sr_2RuO_4$ is also in proximity to a commensurate magnetic tendency that can be stabilized via Fe doping.

Understanding the unconventional superconductivity in high-$T_c$ cuprates, heavy fermions as well as the more recently discovered iron pnictides and chalcogenides, has been attracting tremendous efforts in the past three decades [1-4]. The Ruddlesden-Popper type single-layer ruthenate $Sr_2RuO_4$, an unconventional superconductor [5] with the superconductivity proposed to be chiral $p$-wave that is different from the $s$-wave superconductivity in conventional superconductors or the $d$-wave spin-singlet one in high-$T_c$ cuprates [6], has led to intense theoretical and experimental investigations. Although a variety of experiments have substantiated the unconventional character of the superconducting state and examined the symmetry of the order parameter as well as the structure of the superconducting gap [7-9], the pairing mechanism and the nature of the superconductivity in $Sr_2RuO_4$ are still open questions. For instance, the absence of topological protected edge current [10] is not in line with the time-reversal symmetry-breaking $p$-wave superconductivity [11,12]; recently it is argued that the superconducting Cooper pairs in $Sr_2RuO_4$ cannot be described in terms of pure singlets or triplets, but are spin-orbit entangled states due to spin-orbit coupling [13,14].

Furthermore, as in other unconventional superconductors, the correlation between superconductivity and magnetism in $Sr_2RuO_4$ is of particular interest. That is, the superconductivity is close to magnetic instabilities, and spin fluctuations may be responsible for the superconducting pairing mechanism [15]. While the normal state of $Sr_2RuO_4$ shows Fermi liquid behavior below $T = 25$ K [16], the system exhibits strong magnetic instabilities with ferromagnetic and antiferromagnetic fluctuations coexisting and competing [17,18]: the Fermi surface nesting of the quasi-one-dimensional $\alpha/\beta$ bands (Ru $d_{xz}$, $d_{yz}$) leads to antiferromagnetic fluctuations, while the close proximity of the Fermi level of the quasi-two-dimensional $\gamma$ band (Ru $d_{xy}$) to a Van Hove singularity gives rise to ferromagnetic fluctuations [19,20]. Ferromagnetic correlations have been corroborated by nuclear magnetic resonance (NMR) measurements [21], and are suggested to be responsible for the $p$-wave superconductivity [22]. However, neutron scattering experiments found prominent incommensurate antiferromagnetic fluctuations at $\mathbf{q}_{ic}$ = (0.3 0.3 $L$) (in units of the reciprocal lattice vectors $2\pi/a = 2\pi/b$ and $2\pi/c$), arising from Fermi surface nesting of the $\alpha/\beta$ bands [17]. Such incommensurate antiferromagnetic fluctuations along with strong anisotropy are proposed to account for the unconventional superconductivity in $Sr_2RuO_4$ [23]. Additionally, recent theoretical and experimental studies have also suggested that the superconductivity in $Sr_2RuO_4$ may be generated by the Cooper pairs on the $\alpha/\beta$ bands but not on the $\gamma$ band [24,25].

A fundamental challenge to the understanding of unconventional superconductivity is how the tendency towards magnetic ordering is suppressed while strong magnetic fluctuations are maintained that may lead to superconductivity. Intriguingly, for $Sr_2RuO_4$, at the bare density functional level the incommensurate magnetic instability at $\mathbf{q}_{ic}$ is sufficiently strong so that ordering would be expected [26]. This ordering is presumably suppressed by spin fluctuations, possibly associated with competing orders [18], which is a characteristic common to unconventional superconductors. A powerful means of exploring competing magnetic tendencies in $Sr_2RuO_4$ is chemical doping. For instance, moderate substitutions of Ca for Sr sites, and Ti or Mn for Ru sites, have been shown to give rise to static spin density wave ordering with the same propagation vector as the Fermi surface nesting vector [27-29]. In contrast, carrier doping via La substitution for Sr sites enhances ferromagnetic fluctuations by elevating the Fermi surface closer to the Van Hove singularity of the $\gamma$ band [30]. These studies attest that the magnetic ground state of $Sr_2RuO_4$ is in the vicinity of the antiferromagnetic and ferromagnetic ordering.

In this paper, we report a commensurate, quasi-two-dimensional spin density wave ordering in $Sr_2RuO_4$ induced by Fe substitution for Ru. This magnetic ordered state is characterized by a wave vector $\mathbf{q}_c = (0.25\ 0.25\ 0)$, in contrast to the incommensurate ones in Ti- and Mn-doped compounds [28,29]. Intriguingly, we find that the incommensurate magnetic excitations at $\mathbf{q}_{ic} = (0.3\ 0.3\ 0)$ associated with Fermi surface nesting in pristine $Sr_2RuO_4$ persist in the Fe-doped compounds. This suggests that the induced static ordered state is not driven by Fermi surface nesting, which has been corroborated by our *ab initio* electronic structure calculations. These results imply that, in addition to the known incommensurate magnetic instability, $Sr_2RuO_4$ is also in proximity to a commensurate magnetic tendency which may facilitate the suppression of static magnetic order and give rise to unconventional superconductivity.

The main panel of Fig. 1(a) shows dc magnetic susceptibility $\chi_c$ of $Sr_2Ru_{1-x}Fe_xO_4$ ($x = 0.05$) as a function of temperature measured with $H = 1$ T applied along the $c$ axis. There are three remarkable features. (i) Compared to the weak temperature dependence of magnetic susceptibility associated with the Pauli paramagnetism observed in the parent compound [16], the Fe-doped compound exhibits enhanced Curie-Weiss susceptibility, which implies the existence of localized moment induced by Fe doping. The Curie-Weiss fit of the susceptibility at elevated temperatures gives the effective magnetic moment $\mu_{eff} \sim 1.8\ \mu_B$/Ru. (ii) A paramagnetic-antiferromagnetic phase transition is observed at $T_N \sim 64$ K, as evidenced by the appearance of a peak in the magnetic susceptibility data. (iii) Upon further cooling, a bifurcation between zero-field-cooled (ZFC) and field-cooled (FC) data emerges below $T_g \sim 16$ K, characteristic of a spin-glass-like state. The inset of Fig. 1(a) shows isothermal magnetization measurements performed at $T = 2$ K and 20 K. Hysteresis is observed at 2 K which is consistent with the fact that ferromagnetic correlations develop in the spin-glass-like state. The spin-glass-like state below $T_g$ is also supported by the frequency dependence of ac magnetic susceptibility measurements plotted in Fig. 1(b), where one can see that the peak around 16 K weakly shifts to higher temperature with the increase of the measurement frequency. Note that such a bifurcation between FC and ZFC and the hysteretic behavior in magnetization are absent for the in-plane magnetic susceptibility measurements where the antiferromagnetic phase transition is also observed, as shown in Fig. S1 in the Supplemental Material [32], indicating that the spin-glass-like state presumably arises from the development of short-range ferromagnetic correlations between $RuO_2$ layers. Furthermore, the magnetic moments induced by Fe doping exhibit magnetic anisotropy with the ordered moment along the $c$ axis. Similar features have been observed in the Ti- and Mn-doped $Sr_2RuO_4$ [29,43].

Figure 1(c) presents the temperature dependence of specific heat measured at zero field. An anomaly is observed around $T_N$, corresponding to the onset of antiferromagnetic ordering. The small change in specific heat at $T_N$ might be due to the small magnetic moment size associated with this spin ordered state. It is worth noting that a specific heat anomaly is not convincingly observable in the Ti- and Mn-doped compounds, even though a static magnetic order develops at low temperature in both systems [29,44]. The inset of Fig. 1(c) shows the plot of $C_p / T$ vs $T^2$ and the extracted Sommerfeld coefficient is in the range of $27 - 35$ mJ mol$^{-1}$ K$^{-2}$, depending on the temperature fitting regime, and is slightly smaller than the one obtained for the parent compound [16], presumably due to the reduced carrier density upon the formation of the spin density wave order (more discussions later) [29,44]. Interestingly, as seen in the inset, the specific heat at lower temperature is enhanced and can be suppressed upon applying a 9-T magnetic field, which is most probably ascribable to the magnetic contribution associated with the spin-glass-like state. Temperature dependence of out-of-plane and in-plane resistivity, $\rho_c$ and $\rho_{ab}$, are shown in Fig. 1(d).

Both $\rho_c$ and $\rho_{ab}$ exhibit anomalies at $T_N$ and close to $T_g$. Particularly, the increase in $\rho_{ab}$ below $T_N$ implies partial gap opening of the Fermi surface arising from the onset of antiferromagnetic order.

In order to determine the magnetic structure in $Sr_2RuO_4$ induced by Fe doping, we performed neutron diffraction measurements. Figure 2(a) shows the scans along the [1 1 0] direction over $\mathbf{q}_c$ = (0.25 0.25 0) at $T$ = 4, 50, and 100 K measured on $Sr_2(Ru_{0.95}Fe_{0.05})O_4$. A Gaussian-shaped Bragg peak is clearly observed at 4 and 50K but vanishes at 100K, indicating the magnetic origin of this peak. In addition, the full width at half maximum (FWHM) is found to be determined by the instrumental resolution, which implies the formation of a long-range commensurate magnetic order in the basal plane. On the contrary, the scans around $\mathbf{q}_{ic}$ = (0.3 0.3 0) and (0.3 0.3 1) do not give discernible magnetic diffraction intensity. Figure 2(b) shows the scans along the [0 0 1] direction across the magnetic Bragg peak $\mathbf{q}_c$ = (0.25 0.25 0) measured at various temperatures. Distinct from the scans along the [1 1 0] direction shown in Fig. 2(a), these curves can be well fitted using a Lorentzian function implying a correlation length of ~20 Å along the $c$ axis at $T$ = 4 K. This suggests that the magnetic ordering induced by the Fe substitution in $Sr_2RuO_4$ is nearly two-dimensional, with very short-range magnetic correlation between the $RuO_2$ layers. Additionally, the strongest magnetic Bragg peak observed at $\mathbf{q}_c$ = (0.25 0.25 $L$) with $L$ = 0 instead of $L$ = 1 indicates the absence of the phase shift between neighboring $RuO_2$ layers [27]. These results are in sharp contrast to the earlier studies on Ti- and Mn-doped $Sr_2RuO_4$, where short-range incommensurate spin density wave orderings with the propagation vector $\mathbf{q}_{ic}$ = (0.3 0.3 1) associated with the Fermi surface nesting are reported [28,29], suggesting a different mechanism for the emergence of the commensurate magnetic ordering in Fe-doped $Sr_2RuO_4$. The temperature dependence of the magnetic scattering peak intensity at $\mathbf{q}_c$, which is proportional to the square of the staggered magnetization of the antiferromagnetic order, is shown in Fig. 2(c). A well-defined phase transition is readily seen at $T_N \sim 64$ K, consistent with the magnetic susceptibility and specific heat measurements shown in Fig. 1. It is worth noting that for the 3% Fe-doped compound the magnetic scattering signals are also observed at $\mathbf{q}_c$ and other equivalent positions, but not at $\mathbf{q}_{ic}$, as presented in the contour map in Fig. 2(d). The observation of magnetic reflections associated with the magnetic propagation vectors $\mathbf{q}_c$ = (0.25 0.25 0) and (0.25 -0.25 0) implies the existence of magnetic twin domains due to the tetragonal symmetry of the crystal structure. The intensity of the corresponding magnetic reflections is comparable, indicating that the population of these two magnetic twin domains (Fig. S2) is nearly equal [31].

Possible models of the magnetic structure have been explored by representation analysis using the program *BasIreps* in the FullProf Suite [45], and by the magnetic symmetry approach using the tools available at the Bilbao Crystallographic Server [46]. The maximal magnetic space groups compatible with the parent space group *I4/mmm* and the wave vector $\mathbf{q}_c$ = (0.25 0.25 0) require magnetic moments oriented either along the *c* axis or lying in the *ab* plane. We found that our data are best described by the spin density wave models in which the moments are parallel to the *c* axis, in agreement with the magnetic susceptibility measurement discussed above. Since the moment distribution can be described as a cosine modulation $\mu_l = S\cos(\boldsymbol{qR}_l + \varphi)$, there are two possible spin configurations that depend on the choice of the initial phase $\varphi$: (i) S (+, 0, −, 0) when $\varphi = 0$ (magnetic group $C_cmcm$), or (ii) $1/\sqrt{2}$S (+, +, −, −) when $\varphi = (2n+1)\pi/4$, in which *n* is an integer (magnetic group $C_cmca$). The S represents the amplitude of the magnetic moment which has been estimated from diffraction data to be about ~0.4 $\mu_B$. The schematics of these two magnetic structure models are illustrated in Figs. 3(a) and 3(b), respectively. Note that these two models give rise to identical neutron diffraction patterns and are different only in the local moment size by a factor of $\sqrt{2}$.

The fact that the commensurate magnetic order with a propagation vector of $\mathbf{q}_c = (0.25\ 0.25\ 0)$ emerges in the Fe-doped $Sr_2RuO_4$ is very intriguing, considering that both the strong magnetic fluctuations in the pristine compound and the static incommensurate magnetic order in the Ti- and Mn-doped compounds occur at the same wave vector of $(0.3\ 0.3\ L)$, which is ascribed to the Fermi surface nesting of the quasi-one-dimensional $\alpha/\beta$ bands [17,28,29]. This raises an important question: does the commensurate magnetic ordering originate from the change of nesting vector of the Fermi surface upon Fe substitution? To address this question, we performed density functional theory (DFT) calculations for the pristine and Fe-doped $Sr_2RuO_4$ [31]. The Fermi surfaces and other properties of bulk $Sr_2RuO_4$ were similar to prior reports [22]. All calculations with Fe spin polarized were performed. The density of states (DOS) and projections of a $3 \times 3 \times 1$ supercell, which contains one Fe atom replacing a Ru on $Sr_2RuO_4$, is shown in Fig. 4 along with a band structure plot for the folded zone. The majority spin of Fe $d$ orbitals is fully occupied, as shown in Fig. 4(a), suggesting that Fe enters as high spin configuration $Fe^{3+}$ which is in agreement with the XAS measurements presented in Fig. 3(c). The calculated multiplet splitting of the Fe $3s$ core level in our DFT calculation is 4.45 eV, consistent with this high-spin state. Thus, the introduction of Fe results in an electron deficiency of 1 $e$ / Fe for the host lattice. It is important to note that the Fermi surfaces are large and, by Luttinger's theorem, changes of $0.03\ e - 0.05\ e$ per cell mean changes in Fermi surface volume of $0.015 - 0.025$ of the Brillouin zone volume, consistent with the small shifts (~0.1 eV near $E_F$) along with distortions that we find in the band structure for 11% Fe [Fig. 4(b)-4(d)]. These small changes resulting from 3% and 5% Fe doping then cannot explain the large shift in the magnetic ordering vector we find, and thus a simple itinerant electron explanation in terms of band filling is not operative. However, in addition to the Fe moments, we find a strong back-polarization of the Ru neighboring Fe amounting to more than 1 $\mu_B$/Ru neighbor (1.08 $\mu_B$ as obtained by integration of the spin density over a sphere of radius 2 Bohr around the Ru). We infer that this strong local magnetic coupling of Fe and Ru frustrates the incommensurate nesting and leads to the commensurate order observed in our experiments.

The robustness of the nesting vector of the α/β bands on the Fermi surface with respect to Fe doping is corroborated experimentally by the magnetic excitation spectra measured using the time-of-flight inelastic neutron scattering technique. The lower panel of Fig. 3(d) shows the contour map of the scattering intensity of $Sr_2Ru_{1-x}Fe_xO_4$ ($x = 0.03$) as a function of $E$ (i.e., energy transfer) and $K$. Surprisingly, the dominant magnetic excitations above $E = 3$ meV are well centered at incommensurate positions of $\mathbf{q}_{ic} = (0.3\ 0.3\ 0)$ and $(0.3\ 0.7\ 0)$ [black curve in the upper panel of Fig. 3(d)], which is different from the wave vectors of the elastic magnetic reflections (red curve) centered at commensurate positions $\mathbf{q}_c = (0.25\ 0.25\ 0)$ and $(0.25\ 0.75\ 0)$ [also shown in Fig. 2(d)]. In addition, the magnetic fluctuations barely show any energy dependence, similar to that observed in both the pristine and the Ti-doped compounds [17,47]. While the magnetic excitation related to this ordered state warrants further investigation, the coexistence of the commensurate magnetic order at $\mathbf{q}_c = (0.25\ 0.25\ 0)$ and the dynamic spin fluctuation at $\mathbf{q}_{ic} = (0.3\ 0.3\ 0)$ in the Fe-doped compound implies that the magnetic order is not driven by the Fermi surface nesting as observed in the Ti- and Mn-doped ones [28,29]. Thus, Fe doping reveals a previously unanticipated commensurate magnetic instability in $Sr_2RuO_4$ at $\mathbf{q}_c = (0.25\ 0.25\ 0)$, which competes with the known incommensurate tendency. These results suggest that the tendency towards magnetic ordering in $Sr_2RuO_4$ is suppressed by quantum fluctuations associated with competing magnetic instabilities, while strong spin fluctuations are maintained and may give rise to the unconventional superconducting state.

In summary, we have unraveled a commensurate spin density wave order with a propagation wave vector $\mathbf{q}_c = (0.25\ 0.25\ 0)$ in $Sr_2RuO_4$ upon Fe doping into Ru sites while the incommensurate magnetic fluctuations at $\mathbf{q}_{ic} = (0.3\ 0.3\ L)$ observed in the pristine compound persist. This suggests that this commensurate ordered state does not arise from Fermi surface nesting, in contrast to the previous studies on Ti-, Mn-, and Ca-doped compounds [27-29]. Furthermore, this study indicates that the unconventional superconducting state in $Sr_2RuO_4$ is not only adjacent to the known incommensurate magnetic order but also to a commensurate one.

Work at Michigan State University was supported by the National Science Foundation under Award No. DMR-1608752 and the start-up funds from Michigan State University. Work at Tulane University was supported by the U.S. Department of Energy (DOE) under EPSCOR Grant No. DE-SC0012432 with additional support from the Louisiana Board of Regents (support for crystal growth). Work at ORNL's SNS and HFIR was supported by the Scientific User Facilities Division, Office of Basic Energy Sciences, DOE. This research used resources of the Advanced Photon Source, a U.S. Department of Energy Office of Science User Facility operated for the DOE Office of Science by Argonne National Laboratory under Contract No. DE-AC02-06CH11357.

**Figure Captions**

**Figure 1.** (a) Temperature dependence of out-of-plane dc susceptibility $\chi_c$ of $Sr_2Ru_{1-x}Fe_xO_4$ ($x =$ 0.05). ZFC denotes zero-field-cooled data and FC represents field-cooled data with 1 T measurement field. Inset shows the isothermal magnetization as a function of field measured at 2 and 20 K after ZFC. (b) Temperature dependence of ac susceptibility measured with $h = 10$ Oe. (c) Temperature dependence of specific heat measured at zero field. Inset shows the expanded view of the lower temperature region with the data measured at 9 T included for comparison. The solid red line is the linear fit for 16 K $< T <$ 30 K. (d) In-plane and out-of-plane resistivity as a function of temperature.

**Figure 2.** (a) Neutron diffraction measurement across $\mathbf{q}_c = (0.25\ 0.25\ 0)$ along the [1 1 0] direction at $T = 4$, 50, and 100 K measured on $Sr_2Ru_{1-x}Fe_xO_4$ ($x = 0.05$). (b) Neutron diffraction measurement across $\mathbf{q}_c = (0.25\ 0.25\ 0)$ along the [0 0 1] direction at selected temperatures. (c) The intensity of magnetic Bragg peak $\mathbf{q}_c = (0.25\ 0.25\ 0)$ as a function of temperature. Note that for (b) the sample measured is smaller than that for (a) and (c). (d) Contour map of elastic magnetic scattering intensity of $Sr_2Ru_{1-x}Fe_xO_4$ ($x = 0.03$) at $T = 1.6$ K after subtracting the background measured at 80 K. Spurious peaks are denoted by red circles. The residue intensity near the nuclear peaks ($\pm 1\ \pm 1$ 0) is presumably due to the thermal shift in the lattice parameters.

**Figure 3.** (a),(b) Schematic diagrams of the spin density wave ordering of $Sr_2Ru_{1-x}Fe_xO_4$ ($x = 0.05$). (c) $x$-ray absorption spectra of $Sr_2(Ru_{0.97}Fe_{0.03})O_4$ near the Fe $L$ edge in comparison with FeO and $Fe_2O_3$ indicating the 3+ valence state of Fe dopants. (d) Lower panel: contour map of inelastic neutron scattering intensity as a function of $E$ and $K$, $H$ integrated from 0.2 to 0.4. Upper panel:

the cut along [0 1 0] with the energy transfer $E$ integrated from 3 to 6 meV (black)  and from -0.5 to 0.5 meV (red), respectively. $H$ is integrated from 0.2 to 0.4. Note that the intensities of these two curves are scaled. Data were measured on $Sr_2Ru_{1-x}Fe_xO_4$ ($x = 0.03$).

**Figure 4.** Electronic structure for a $3 \times 3 \times 1$ supercell of $Sr_2RuO_4$ containing one Fe substitution. (a) Density of states and projections, showing majority spin as positive and minority spin as negative, implying that the Fe majority $d$ bands are filled corresponding to $Fe^{3+}$. (b) Fat band plot of the band structure showing Ru character for the unsubstituted supercell (heavier symbols mean higher Ru character), in comparison with the Fe substituted cell, emphasized by heavier symbols. (c) Ru character from Ru neighboring Fe and (d) Ru not neighboring Fe.

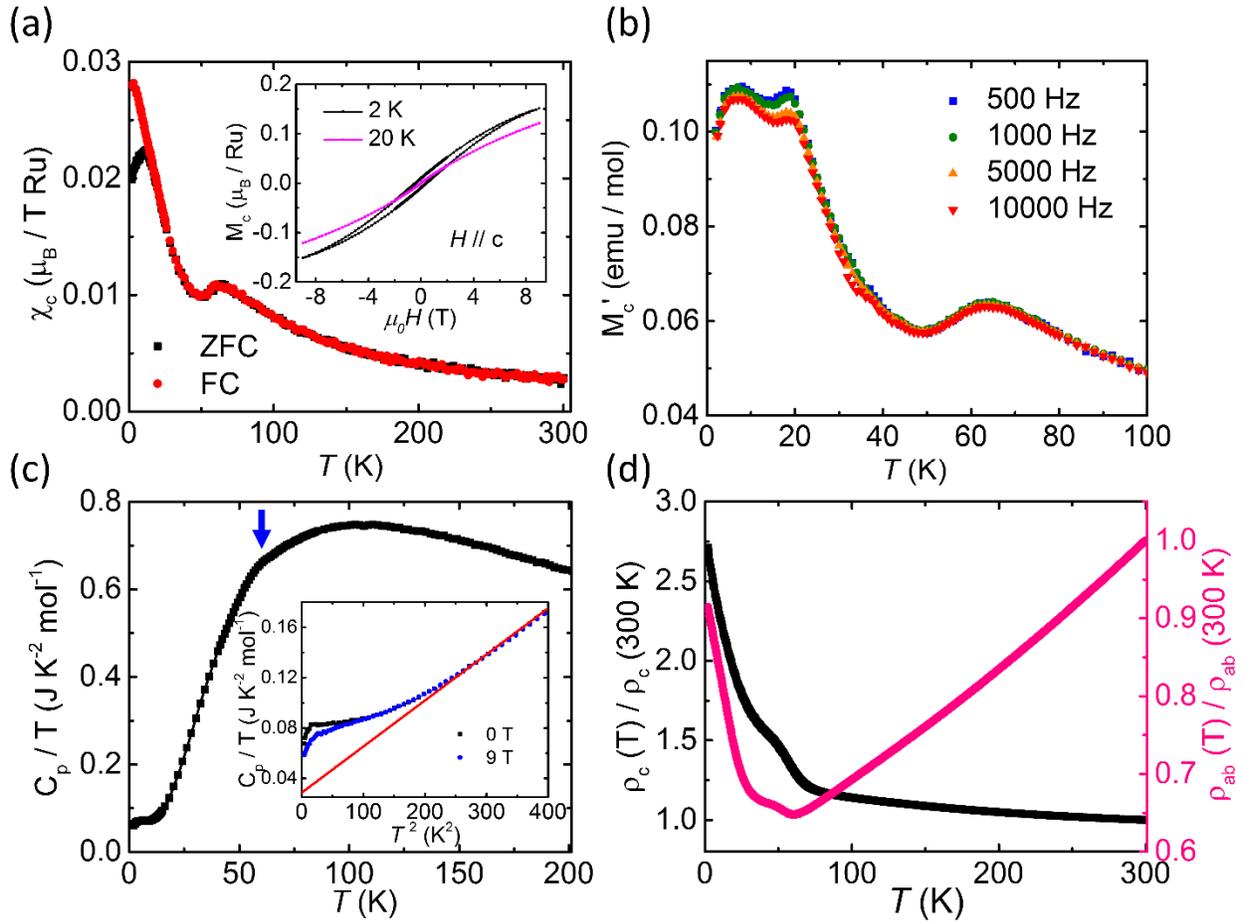

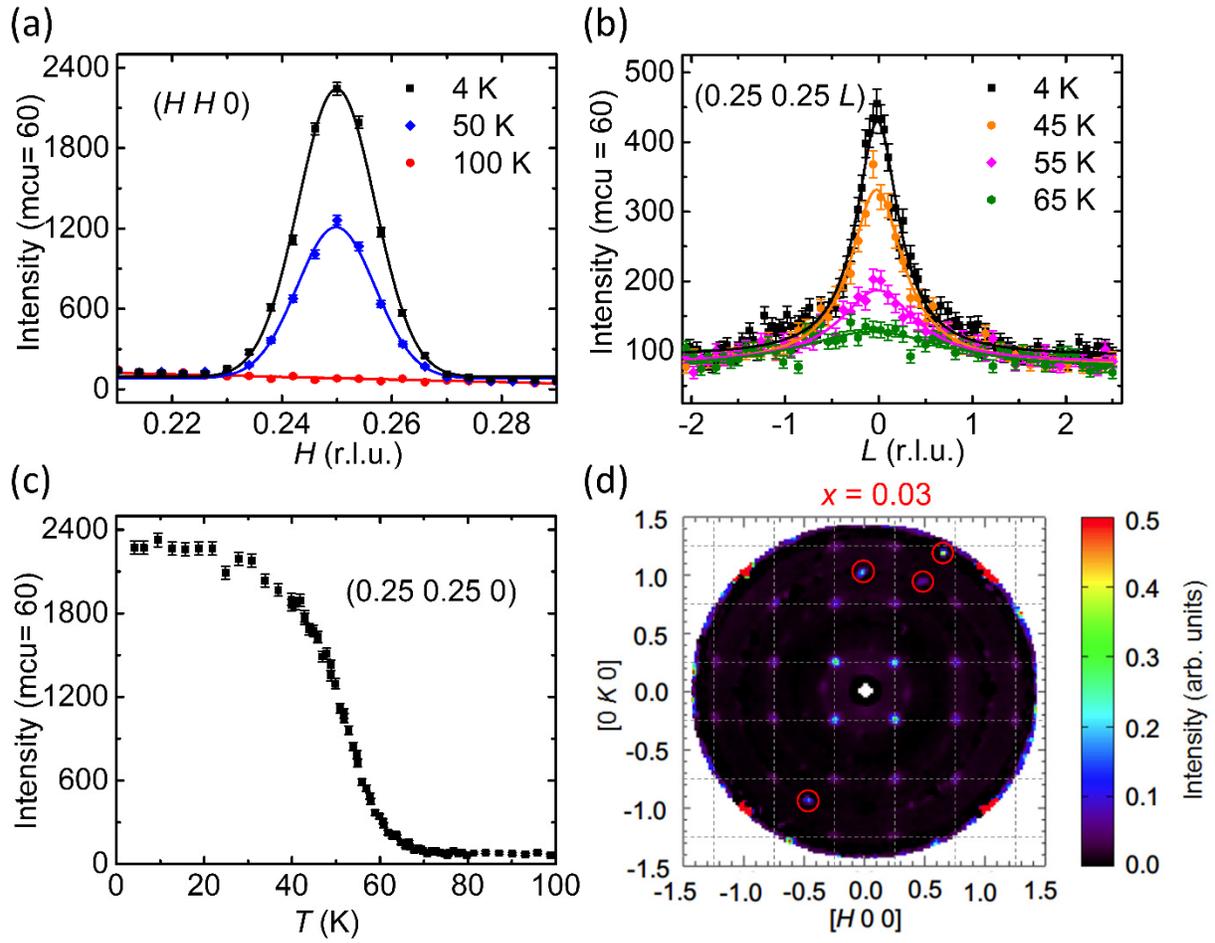

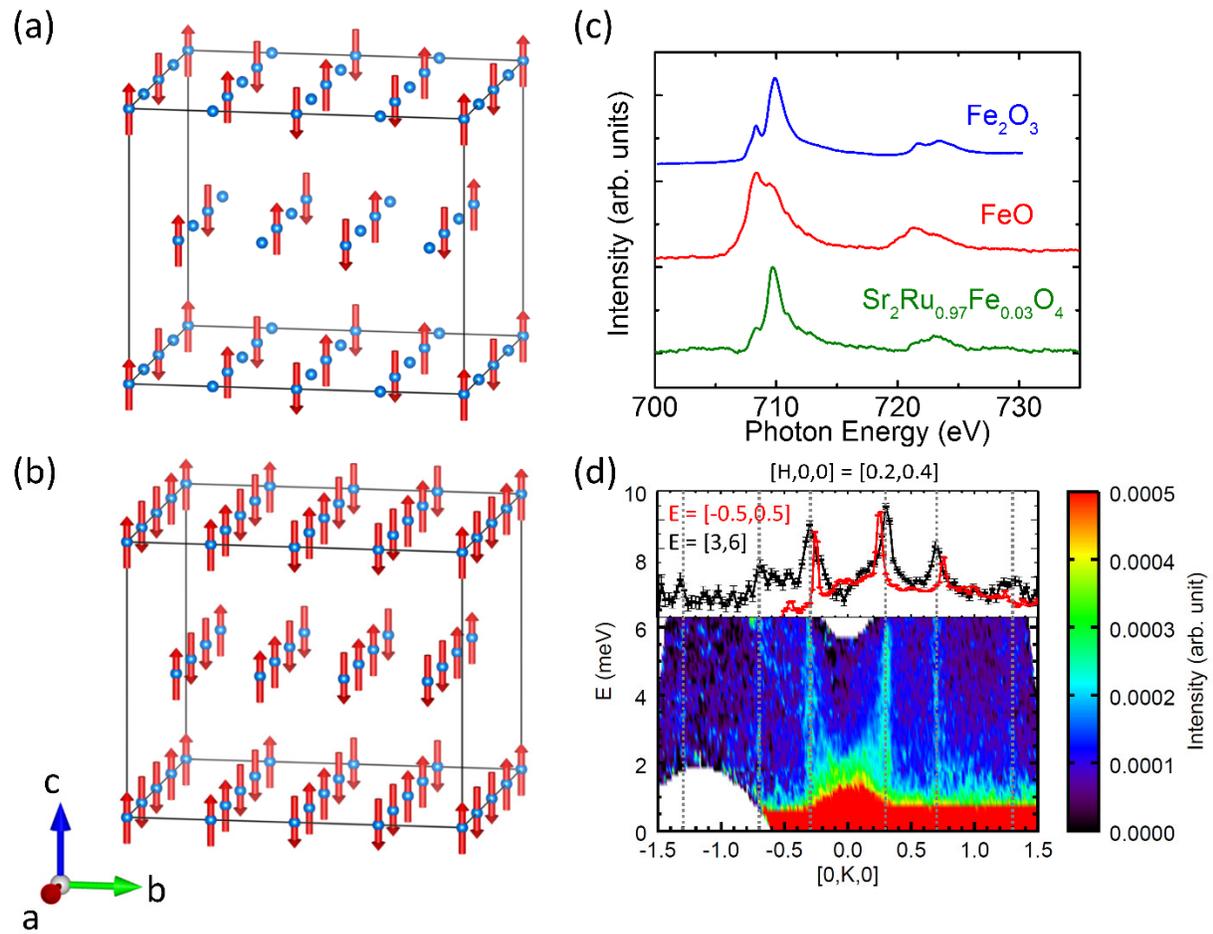

**Figure 4**

M. Zhu *et al.*

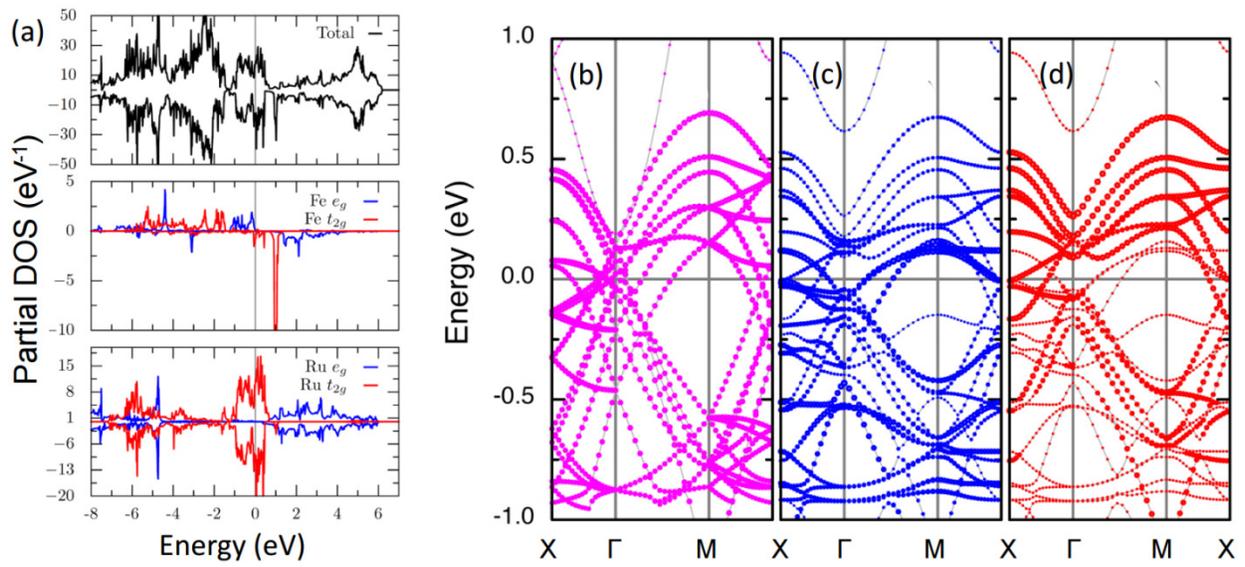